\documentclass{sig-alternate-10pt}
\usepackage{graphicx}
\usepackage{subfigure}
\usepackage{algorithmic}
\usepackage{amssymb}
\usepackage{url}
\usepackage{algorithm}
\newcommand{\eg}{{\it e.g.}}

\newcommand{\etc}{{\it etc.}}

\usepackage{color}
\usepackage{ifthen}
\newboolean{TECHREPORT}
\setboolean{TECHREPORT}{false}
\usepackage{cite}
\usepackage{amssymb,amsmath}
\usepackage{newclude}
\usepackage{epstopdf}

\usepackage[T1]{fontenc}
\usepackage[latin2]{inputenc}
\usepackage{balance}

\definecolor{heraldBlue}{rgb}{0.0,0.0,0.8}
\definecolor{heraldRed}{rgb}{0.8,0.0,0.0}
\definecolor{heraldGray}{rgb}{0.4,0.4,0.4}
\definecolor{heraldBlack}{rgb}{0.0,0.0,0.0} %
\definecolor{heraldGreen}{rgb}{0.0,0.4,0.0} %

\begin{document}

\title{Crowd-assisted Search for Price Discrimination in E-Commerce: First results}

\numberofauthors{1}

\author{
\small 
\alignauthor Jakub Mikians$^\dagger$, L\'aszl\'o Gyarmati$^\star$, Vijay Erramilli$^\star$, Nikolaos Laoutaris$^\star$ \\
       \affaddr{Universitat Politecnica de Catalunya$^\dagger$, $^\star$Telefonica Research}\\
       \email{jmikians@ac.upc.edu,\{laszlo,vijay,nikos\}@tid.es}
}

\maketitle

\setlength{\emergencystretch}{3em}

\begin{abstract}

After years of speculation, price discrimination in e-commerce driven by the personal information 
that users leave (involuntarily) online, has started attracting the attention of privacy researchers, 
regulators, and the press. In our previous work we demonstrated instances of products whose prices varied online depending on the location and the characteristics of perspective online buyers. In an 
effort to scale up our study we have turned to crowd-sourcing. Using a browser extension we have collected 
the prices obtained by an initial set of 340 test users as they surf the web for products of their interest. This 
initial dataset has permitted us to identify a set of online stores where price variation is more pronounced. We have 
focused on this subset, and performed a systematic crawl of their products and logged the prices obtained from different 
vantage points and browser configurations. By analyzing this dataset we see that there exist several retailers that return 
prices for the same product that vary by 10\%-30\% whereas there also exist isolated cases that may vary up to a multiplicative factor, \eg, $\times 2$. To the best of our efforts we could not attribute the observed price gaps to currency, shipping, or taxation differences.

\end{abstract}

\section{Introduction}

Price discrimination refers to the practice of selling the same product to different customers at different prices 
that depend on individual customer's willingness to pay. The system of fixed prices for goods that we are used to
today is mostly a 20th century phenomenon whereas price discrimination is probably as old as commerce itself. 
It has been employed by merchants to extract higher profit margins from customers that are willing to pay more 
for a product while making sure that more price sensitive customers are retained by offering them a lower price. 

With the rise of e-commerce in the last decade many expected prices to move strictly in one direction -- downwards -- 
as a result of more intense competition fueled by the customers' ability to compare online the prices of different 
retailers. It was not long before the first concerns appeared with the conjecture that online shopping could backfire 
for customers in the form of price discrimination driven by the personal information of users 
collected by various online entities~\cite{Odlyzko2003}. Such a possibility would further erode online privacy. For example, 
users frequenting luxury product websites or geo-located to certain ZIP codes could be tagged 
as affluent or price insensitive and consequently be displayed inflated prices.  

We tested this conjecture in a recent paper and were able to demonstrate a few examples in which 
the prices of online offerings seemed to vary (please refer to~\cite{hotnetspd} for concrete examples). 
In order to broaden the scope of our measurements so that we can derive general conclusions regarding
the frequency and magnitude of suspected price discrimination, we turn to crowdsourcing. 
Crowdsourcing enables end-users to (i) point us to products and e-retailers that might be engaging in price discrimination, 
and (ii) aid us in extracting the prices of products from web pages without requiring manual 
intervention (Sec.~\ref{sec:context}). Crowdsourcing, therefore helps us in scaling up the search process. 
This is achieved by a browser extension called \$heriff developed by one of the authors~\cite{Jakub}, (Sec.~\ref{subsec:sheriff}). 
In this paper we present the first results obtained during the beta testing phase of \$heriff that lasted for a three
month period (Sec.~\ref{subsec:crwddata}). These results pointed to price variations observed in well known, 
but also in relatively unpopular sites and categories as well, different from our observations in~\cite{hotnetspd}, 
consistently over time and across different locations,
underscoring the effectiveness of the crowdsourcing approach.
We then perform a systematic measurement study of products on this set of e-retailers by performing a large crawl (Sec.~\ref{sec:deepdive}) and understand the conditions that can lead to price variations 
Our main results include the magnitude of price variations for most e-retailers is between 10\%--30\%, the cheapest
products often face the highest variation ($\times 3$) with the most expensive ones having lower variation ($\times 1.5$)
, and physical location plays a role in price variations for different categories of products. 

\section{Setting the context}
\label{sec:context}
In this section, we set the context for our study by first discussing the questions we tackle,
the challenges in answering these questions, and how we address them. 

\subsection{Open questions}

\begin{itemize}

\item Do we see persistent, reproducible price variations and which e-retailers engage in price variation? 

\item How frequent and large are the observed variations? Which products experience 
price variations (cheaper or more expensive ones) and what type of 
variation (additive/multiplicative) do we see? 

\item Can we attribute price variation to actual price discrimination? In general, it is impossible to 
assert without access to the code that generates the prices that any price variation we observe is in reality price discrimination. However, we can eliminate several alternative causes that might explain them as discussed later. 

\item Finally, when there are price variations, can we attribute them  to specific personal information traits (location, browsing history, \etc)?

\end{itemize}

\subsection{Challenges}

Any system wanting to perform large scale search for price discrimination has to parse product pages, extract the location of the price from web pages, and fan out queries to the same product page from other vantage points in order to compare the results.  

The challenges that need to be addressed
are as follows: (i) Different retailers have different web templates for presenting their products. Extracting the price 
of a product from an unknown template is non-trivial: a simple search for dollar or euro sign would fail since typically 
product pages include additional recommended or advertised products along with their prices. Therefore, for each retailer 
one needs to understand its template format and then write a specialized script for 
extracting the price. %
The problem with this is that it cannot scale with the number of retailers. 
(ii) Minimize noise as well as other possible reasons for price variations. Sources of noise include the 
retailer conducting A/B testing, timing difference between original and additional requests for comparison, and 
pricing format differences (different currencies, \etc). There are also reasons like taxation, logistics, shipping costs, intellectual property issues that can cause price differences that are not due to discrimination. For proper attribution of price 
discrimination, we need to ensure the known reasons cannot explain the variations. (iii) In order to better explain 
price discrimination, we need to control for factors like physical location, system issues, and browsing history. 

\textbf{Addressing challenges}

To address scaling issues, we resorted to crowdsourcing, using \emph{\$heriff} a browser extension for Firefox and Chrome. 
Crowdsourcing enables us to outsource the search for price variations and cover a larger part of the web. We describe the tool
briefly in Sec.~\ref{subsec:sheriff}. The results from the tool uncover e-retailers that engage more in varying prices and this lets us
focus more on these e-retailers, expanding scope and depth. %

We took several steps in order to deal with noise. First of all, we synchronized the measurements from different vantage points so that they occur almost at the same time. This reduces the chance that an observed variation is because of time spread, availability, \etc. Also we repeated the same
set of measurements multiple times to guarantee that the results are repeatable. This decreases the possibility of A/B testing and small-scale temporal effects being the cause of price variations. 

Our different vantage points access always the same retailer site, but can be displayed prices on different currencies (the local 
one) because retailers typically geo-locate their IP address. We convert the prices obtained by the different vantage points for the same product into US dollars using the daily lowest and highest exchange 
rates. We keep only products whose price variation is strictly greater than the maximum gap that can exist given the two extreme exchange rates in our dataset. This guarantees that the observed price differences are not due to currency translation issues.

For factors like taxation, shipping costs, and custom duties, we manually checked
to ensure these reasons cannot explain the price differences. Most e-retailers do not include shipping 
and taxing before checkout thus the great majority of our measurements was not affected by such issues. 
Custom duties are in most cases paid post sale directly between the customer and the custom authority without the intervention of the retailer.

\section{Crowd-sourcing}
In this section, we first describe the tool that was used to enable crowdsourcing and then detail the data we have collected
using the tool. We end this section with an analysis of the collected data, which points to the retailers where
price variations are prevalent, as seen by users around the world. 

\subsection{\$heriff}
\label{subsec:sheriff}

We used a browser extension for Firefox and Chrome called \emph{\$heriff}. The extension performs the following tasks: (i) Enables the user to highlight a price of a product on an e-retailer, (ii) once the
price is highlighted, the extension enables the user to check for price variations via a small click button, (iii) when the
button is clicked, the exact URI is sent to 14 vantage points around the world where the same URI is requested and the
entire webpage is downloaded, (iv) given the user has highlighted the price on the page, we use that information to extract the price 
from the downloaded page at different locations, (v) we send these prices back to the user from various locations. The user,
therefore can observe if there are any variations for the exact product she searched for. Hence, the users have an incentive to return to \emph{\$heriff} time to time to check prices again. (vi) We store the pages for analysis
in a database. The extension can be found at: \texttt{http://pdexperiment.cba.upc.edu/}. 

As can be observed, we cannot control for the physical locations when the original query comes from, nor can we control
for the system and/or the browsing history of the user who originated the query. %

\subsection{Collected data and analysis}
\label{subsec:crwddata}

We use a \emph{crowdsourced dataset} collected by \emph{\$heriff} that contains 1500 requests (between Jan-May 2013) to check 
the prices of different products. The requests were issued by 340 different users from 18 countries. In total, the users 
of \emph{\$heriff} checked products from 600 domains. Afterwards, we systematically crawled the sites of retailers 
where \emph{\$heriff} revealed price differences. Before the analyses, we removed the noise from the crowdsourced dataset. Causes behind the noise include diverse number and date formats across countries, product customization not encoded on the URI, \etc. The \emph{crawled dataset} focuses on 21 retailers. We 
randomly picked up to 100 products per retailer and checked the prices of these products on a daily basis for a week. 
The crawled dataset has 188K extracted prices in aggregate. 

\textbf{Which retailers return dynamic prices?}

\begin{figure}[tb]
\centering
\includegraphics[clip=true,trim=0cm 1cm 0cm 0.5cm,width=9cm]{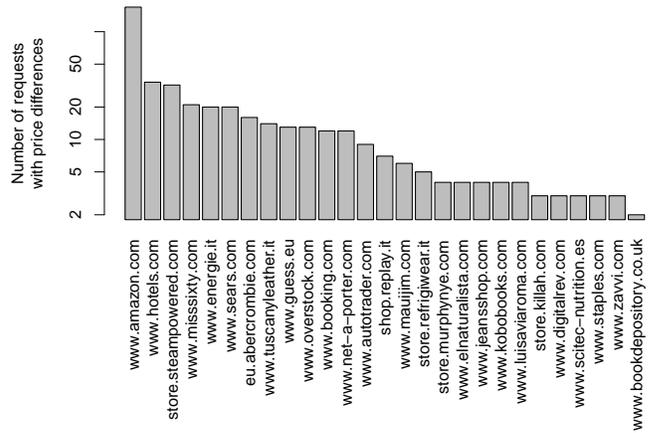}
\vspace{-7mm}
\caption{Domains with the highest number of request where price differences occurred}
\label{fig:result_sheriff_request}
\end{figure}

\begin{figure}[tb]
\centering
\includegraphics[clip=true,trim=0cm 1cm 0cm 0.2cm,width=9cm]{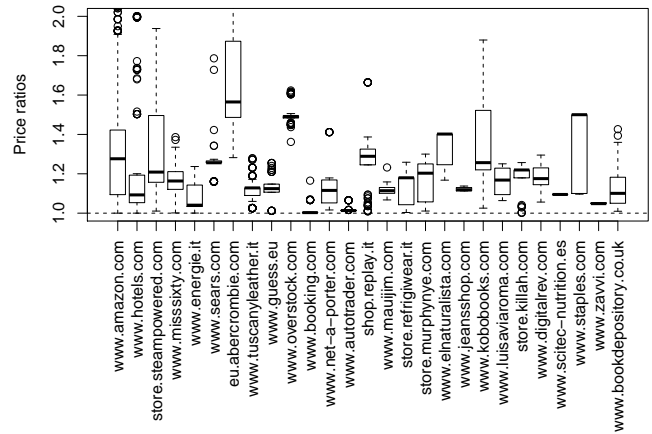}
\vspace{-7mm}
\caption{Magnitude of price differences per domains}
\label{fig:result_sheriff_ratio}
\end{figure}

Fig.~\ref{fig:result_sheriff_request} lists the retailers with the highest number of instances of price variations in the crowdsourced dataset. 
The list includes a diverse set of sites that include bookstores, cloth retailers/manufacturers, office supplies/electronics, 
car dealers, department stores, hotel and travel agencies, \etc.  For each one of these retailers, and for each one
of the products checked on these retailers, 
we computed the ratio between the maximum and minimum price observed across the different measurement points. 
In Fig.~\ref{fig:result_sheriff_ratio} we plot the basic statistics (median, 25-, 75-percentile, and extreme values) of this ratio across all checked products in the dataset for each one of the retailers with the highest frequency of price variation. One can note that a variety of stores return prices that may vary 
anywhere between 15\%-40\% depending on the retailer, whereas there also exist few cases where the difference 
approaches a factor of $\times 2$!  We note here that several of these retailers are not very popular (\texttt{www.elnaturalista.com})
and, in many cases, local (\texttt{store.refrigiwear.it}), underscoring the usefulness of crowdsourcing, as these
retailers were not observed in previous studies~\cite{hotnetspd}.

\section{Crawling specific e-retailers}
\label{sec:deepdive}

\subsection{Retailers}

\begin{figure}[tb]
\centering
\includegraphics[clip=true,trim=0cm 0cm 0cm 0.2cm,width=9cm]{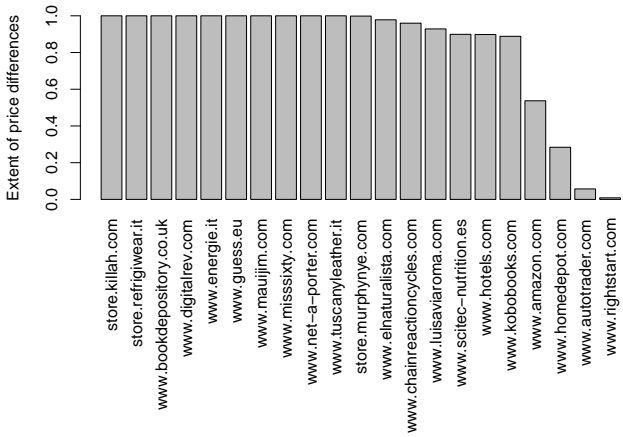}
\vspace{-10mm}
\caption{Measure extent of price variations for different domains}
\label{fig:result_bulk_freq}
\end{figure}

\begin{figure}[tb]
\centering
\includegraphics[width=9cm]{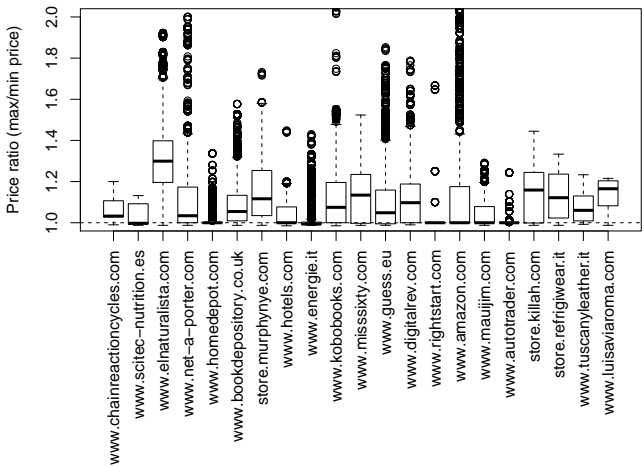}
\vspace{-8mm}
\caption{Magnitude of price variability per domain}
\label{fig:result_bulk_ratio}
\end{figure}

Fig.~\ref{fig:result_bulk_freq} and Fig.~\ref{fig:result_bulk_ratio} depict the same metrics with Fig.~\ref{fig:result_sheriff_request} 
and Fig.~\ref{fig:result_sheriff_ratio} but for the crawled instead of the crowdsourced dataset (Sec.~\ref{subsec:crwddata}). Fig.~\ref{fig:result_bulk_freq} 
shows the fraction of requests we sent out to each retailer that had price variation. In some cases, we see
a 100\% coverage, pointing to the fact
that price variations are a persistent and repeatable phenomenon. 
Indeed, for the majority of retailers in the crawled dataset, we see the extent of price variation 
to be near complete (100\%). 
In terms of the magnitude of price variability, Fig.~\ref{fig:result_bulk_ratio} depicts values 
between 10\% and 30\% for most of the retailers---a non-trivial amount.

\subsection{Looking into products}

\begin{figure}[tb]
\centering
\includegraphics[clip=true,trim=0cm 0.3cm 0cm 1.8cm,width=9cm]{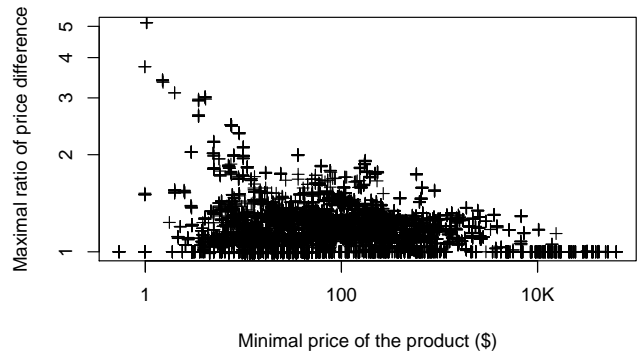}
\vspace{-8mm}
\caption{Maximal ratio of price differences per product price (all stores)}
\label{fig:result_bulk_ratio_price_max}
\end{figure}

We now characterize price variations from the perspective of products. One open question is to understand if there
is any correlation between the price of a product and the magnitude of the price variations associated with that product. 
For each product in crawled dataset (across all retailers) we compute the ratio between the maximum and minimum price across 
our measurement vantage points and plot them in Fig.~\ref{fig:result_bulk_ratio_price_max} against 
the minimum observed price of each product. The figure shows price differences occurring in the entire range of products costing from \$10 to \$10K. The highest differences 
are observed with cheaper products in the order of tenths of dollars, in which case differences up to $\times 3$ are observed. 
We also observe differences up to $\times 2$ for expensive products( in the \$1K range). For the most expensive products 
going into the multiple thousands, the price gap appears to be always smaller than $\times 1.5$.

\begin{figure*}[tb]
\centering
\subfigure[\texttt{www.digitalrev.com}]{\includegraphics[clip=true,trim=0cm 0cm 0cm 2cm,width=2.3in]{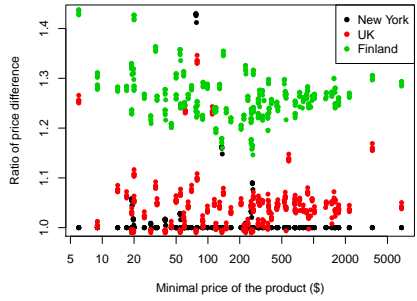}}
\subfigure[\texttt{www.energie.it}]{\includegraphics[clip=true,trim=0cm 0cm 0cm 2cm,width=2.3in]{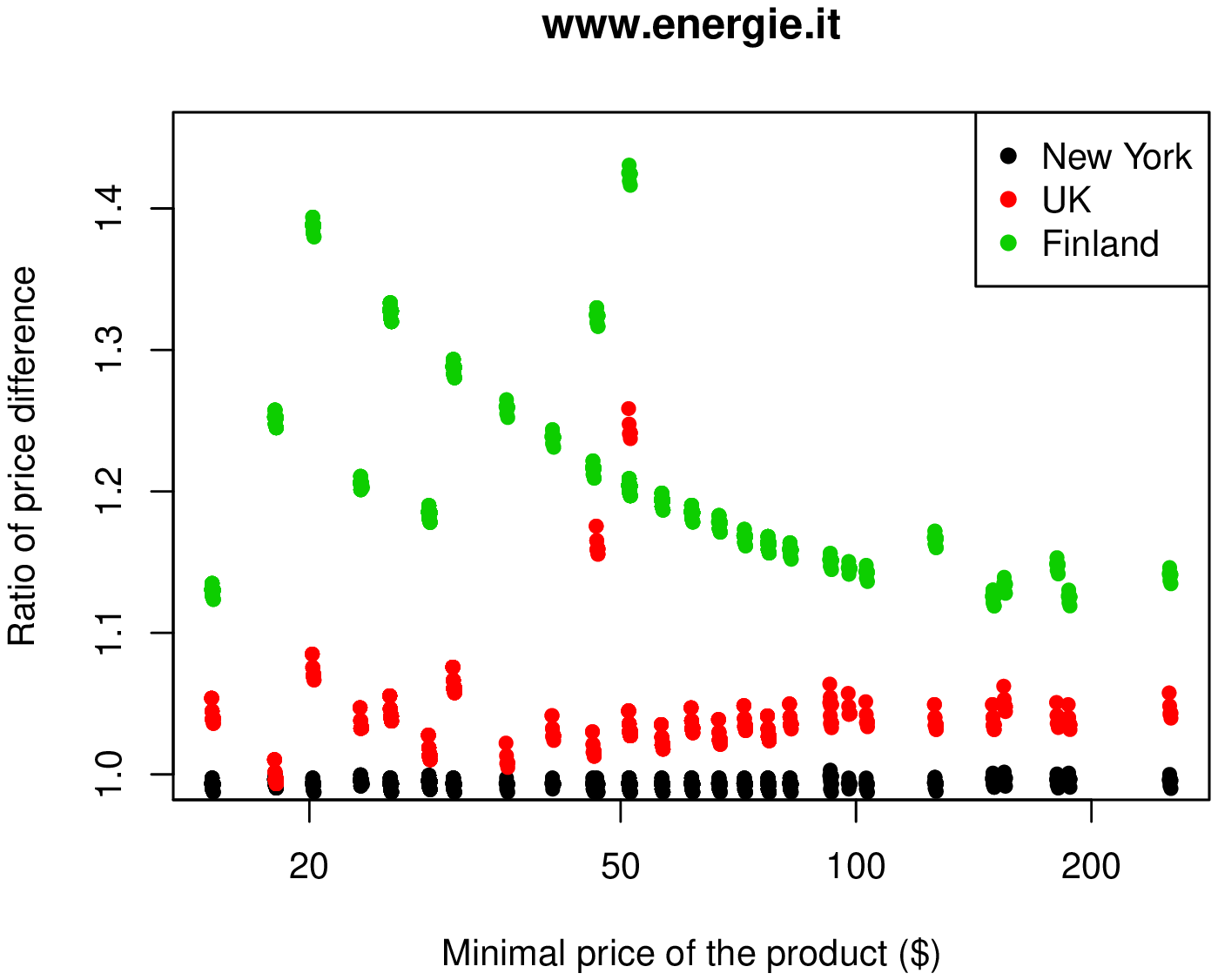}}
\vspace{-5mm}
\caption{Ratio of price differences per product price}
\label{fig:result_ratio}
\end{figure*}

In Fig.~\ref{fig:result_bulk_ratio_price_max} the practices of a diverse set of retailers are mixed together. 
In order to unearth if there are difference strategies that are employed behind varying prices,  
we focus on individual retailers. In Fig.~\ref{fig:result_ratio}(a) we look 
at a retailer of photography equipment. For each one of the products from the retailer we studied, we plot a number of dots 
that is equal to the number of measurement points using different colors to indicate each one of the vantage points. The 
x-axis denotes the minimum price of the product across all locations whereas the y-axis denotes the ratio between the price 
at the location of the dot and the minimum price. One can see parallel (to the x-axis) lines of different colors. 
This in effect means that the price variations between locations is \emph{multiplicative}, equal to the gap 
between two lines on the y-axis, and this applies for the whole range of products (cheap as well as expensive ones). In 
Fig.~\ref{fig:result_ratio}(b) we show the same information from a clothes manufacturer. In this case we see a 
similar behavior for all but one location (green color). In that location the prices vary by an \emph{additive} term 
compared to other locations. As the products become more expensive, the effect of the additive terms 
is progressively eliminated and the lines become parallel from \$100 and onwards. We have other 
examples of retailers that apply a mix of multiplicative and additive pricing across our vantage points. 

\subsection{Does location have an impact?}

\begin{figure}[tb]
\centering
\includegraphics[clip=true,trim=0cm 0.7cm 0cm 0.5cm,width=9cm]{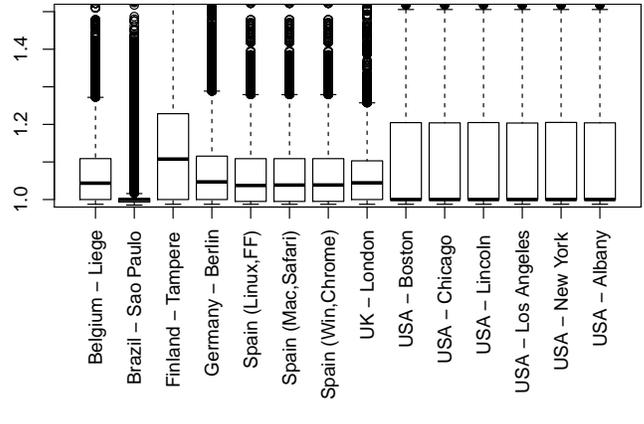}
\caption{Magnitude of price differences per location (all)}
\label{fig:result_allbulk_ratio_location}
\end{figure}

Next we focus our attention on location. At a high level the question that we want to answer is whether 
users from certain locations tend to pay more for the same product than others. As with our previous analysis around 
products, we begin by showing aggregate results across all the retailers we focused on. For each product we 
compute the ratio of its price at a certain location over the minimum price across all locations for the same product. 
In Fig.~\ref{fig:result_allbulk_ratio_location} we present box-plots summarizing the main statistics of the above ratio 
for each one of the locations where we had a measurement vantage point. From a first glance it seems that locations in 
USA and Brazil tend to get lower prices than locations in Europe. Within Europe, Finland stands out as the most expensive location.    

\begin{figure*}[t]
\centering
\subfigure[\texttt{www.homedepot.com}]{\includegraphics[width=2.2in]{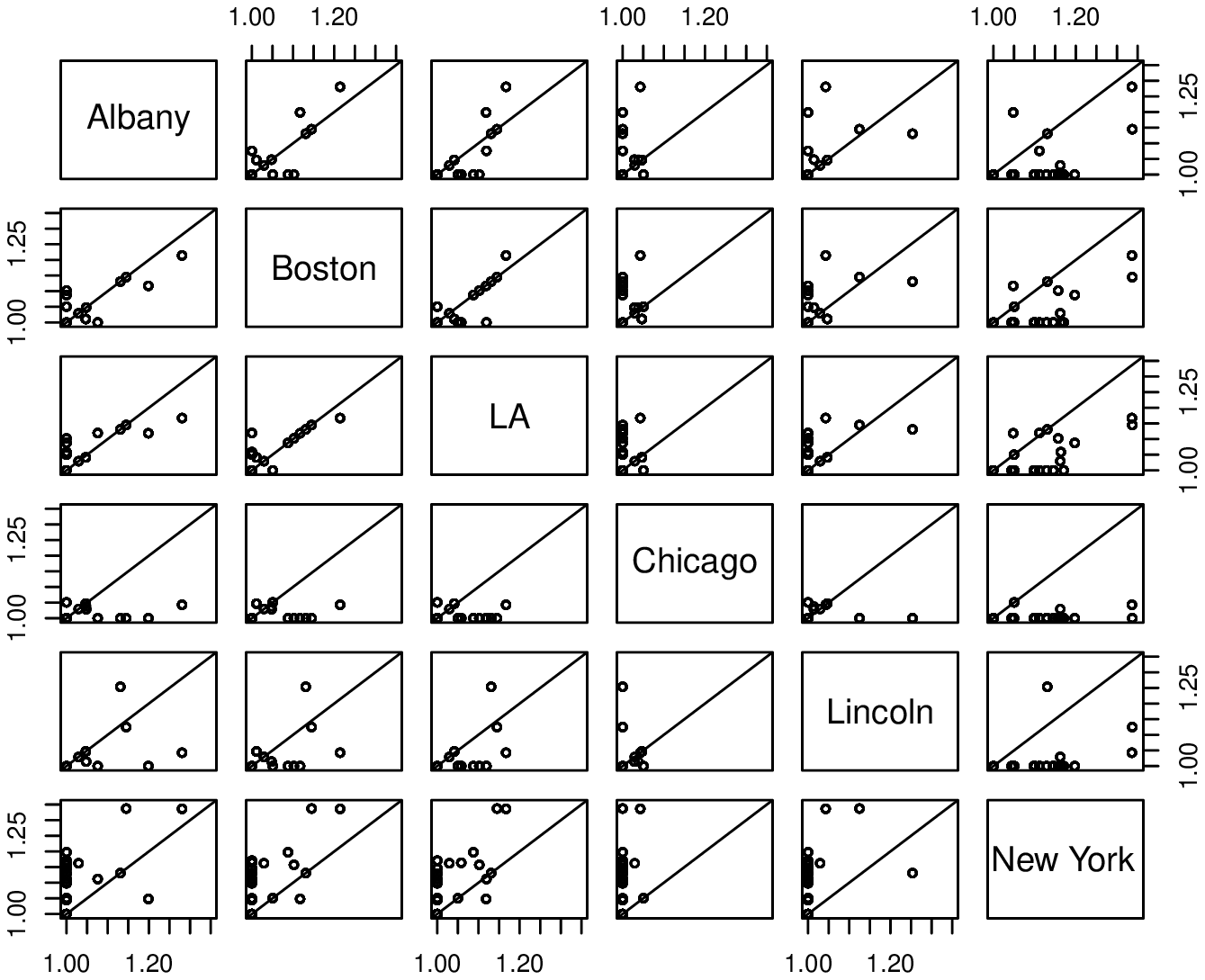}}
\subfigure[\texttt{www.amazon.com}]{\includegraphics[width=2.2in]{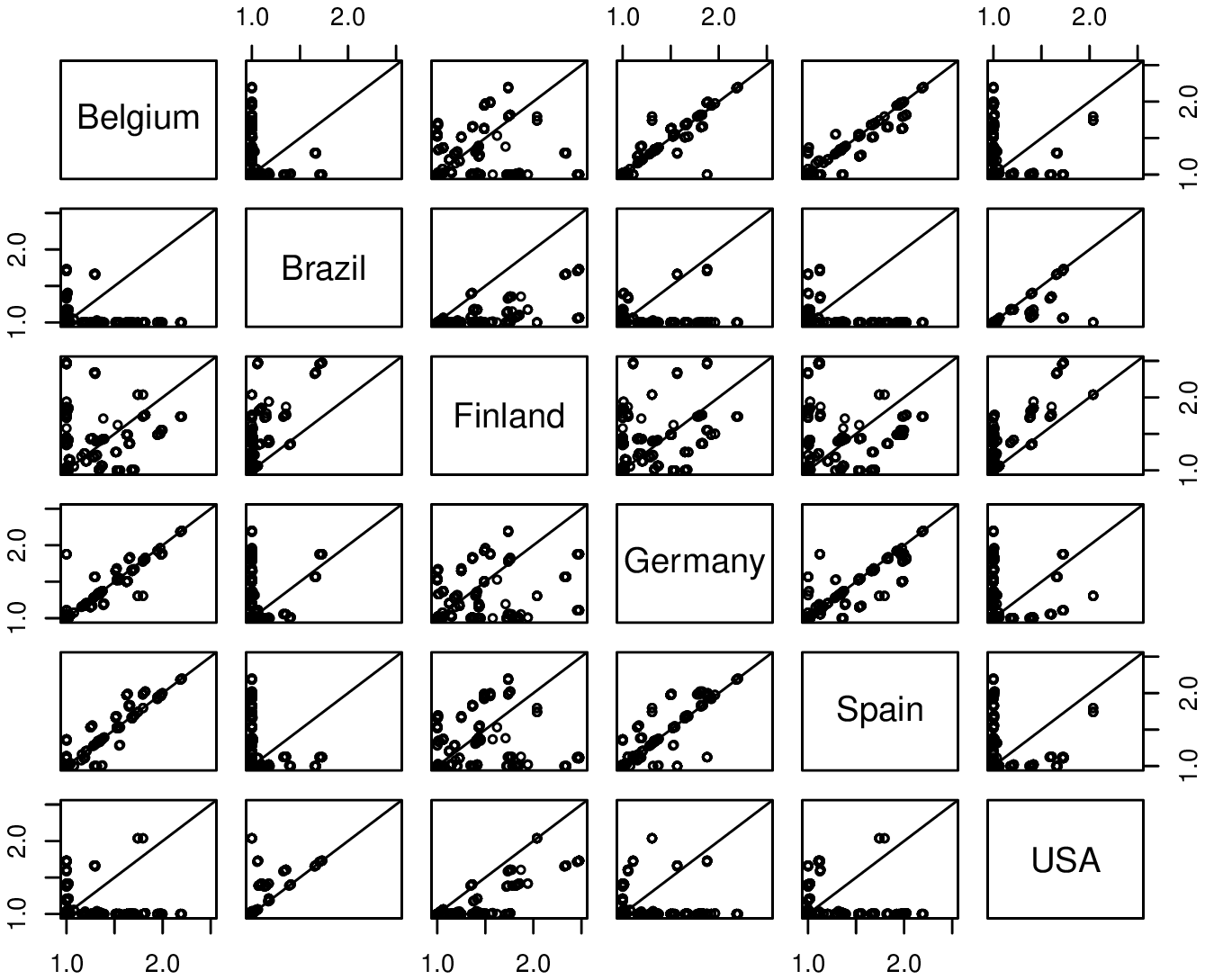}}
\subfigure[\texttt{store.killah.com}]{\includegraphics[width=2.2in]{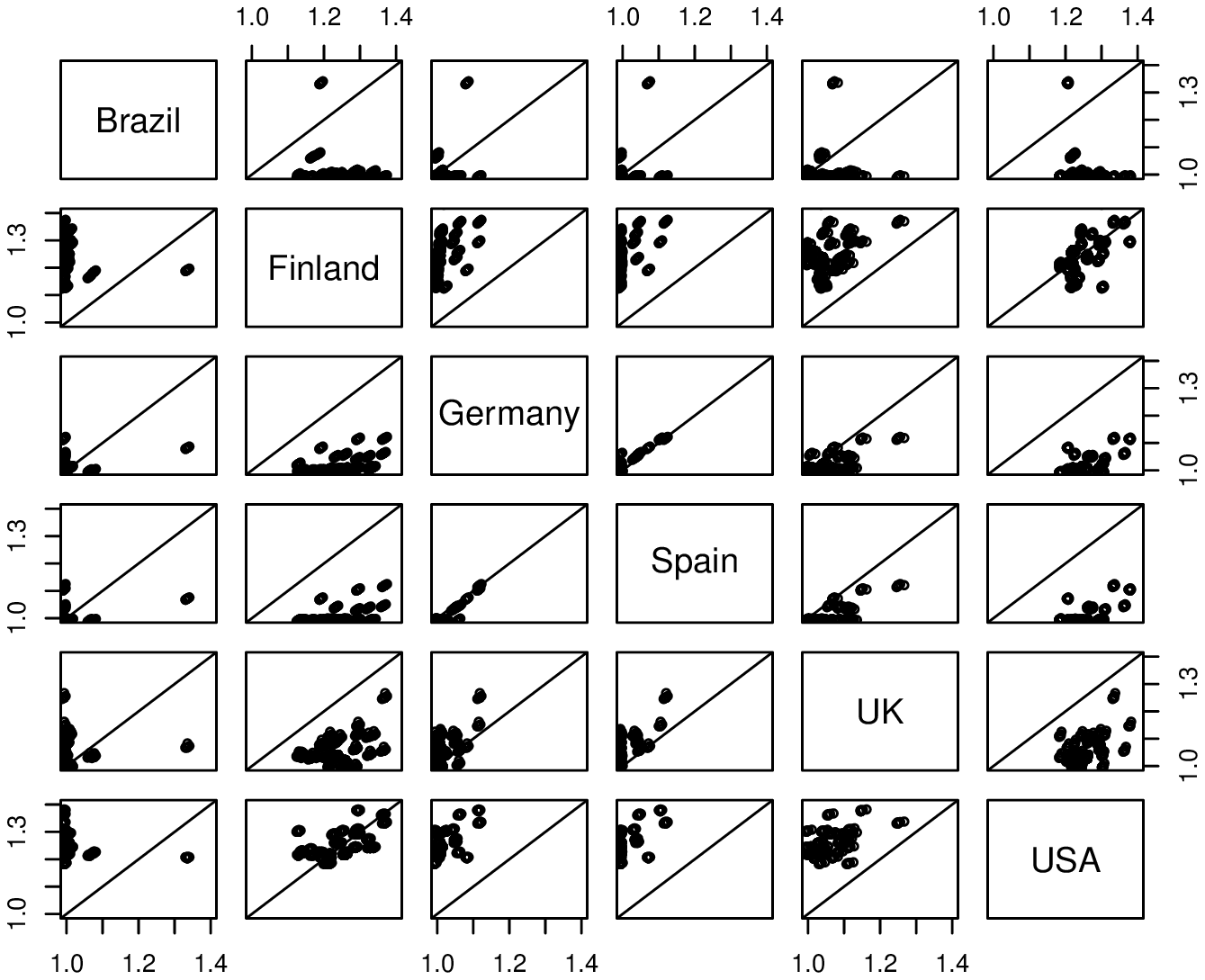}}
\vspace{-5mm}
\caption{Magnitude of price difference per location}
\label{fig:result_ratio_location}
\end{figure*}

To delve deeper into the effect of location we will refine the presented results by (i) focusing on specific retailers, and by (ii) 
presenting pair-wise comparisons of how a retailer prices its products at two different locations. 
We start with a retailer of home improvement appliances and equipments and look at its pricing across 6 US 
cities (Albany, Boston, LA, Chicago, Lincoln, New York). Fig.~\ref{fig:result_ratio_location} (a) presents a 
grid of pairwise comparison subplots. The y-axis for each plot corresponds to the location represented in the row, while the x-axis
for each plot represents the location shown in the column. For example 
subplot(1,2) has Albany on the y-axis and Boston on the x-axis. %
Within a subplot there exist points that correspond to individual products of the said retailer. The y-axis denotes 
the ratio between the price of the product at the y-axis location of the subplot and the minimum price 
of the product across all locations where we have vantage points. The x-axis denotes the same ratio with respect to the 
x-axis location of the subplot. Given these definitions, it is easy to note that a subplot where most of 
the dots fall along the main diagonal of the subplot signifies two locations that get similar prices from the said 
retailer across its products. If the dots cluster closer to the y-axis, then this is a sign that the y-axis location 
is more expensive than the x-axis location and inversely if the dots cluster along the x-axis. 

With the above in mind we can identify a diverse set of pricing relationships. For example, we see that LA and 
Boston (subplot(3,2)) get similar prices, since most of the dots are aligned across the main 
diagonal (similarly with Albany and Boston (subplot(1,2) or (2,1)). On the other hand there exist examples where one 
location observes higher prices than the other -- New York for example, appears to be consistently more expensive 
than Chicago (subplot(6,4)). There also exist mixed cases of pairs where one location is more expensive for some of the 
products but cheaper for some others, \eg, Boston and Lincoln (subplot(2,5)). It is interesting to note that with different 
retailers these relationships change. Also, there exist retailers that have constant prices across US but vary them across countries, 
for example \texttt{amazon.com}, whose pairwise grid is shown in Fig.~\ref{fig:result_ratio_location} (b). A diverse set of 
behaviors include equal price, more expensive/cheaper, and mixed can be observed across different countries. A third example 
from a clothes retailer is depicted in Fig.~\ref{fig:result_ratio_location} (c).

\begin{figure}[tb]
\centering
\includegraphics[clip=true,trim=0cm 0cm 0cm 0.4cm,width=9cm]{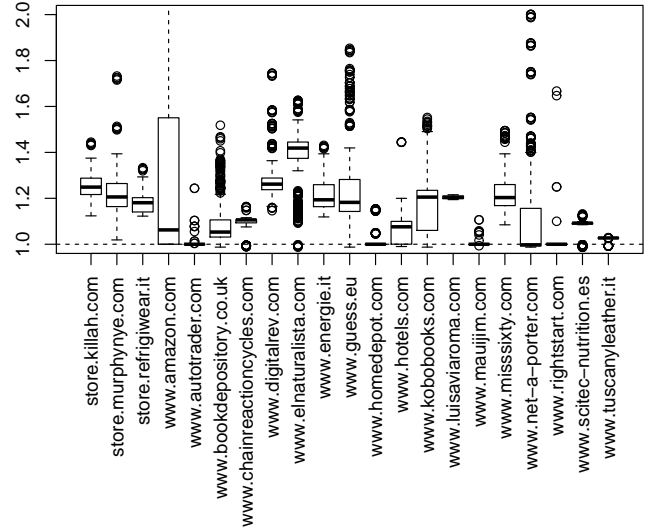}
\vspace{-1cm}
\caption{Magnitude of price differences per domains in Tampere, Finland}
\label{fig:result_Finland_ratio_location}
\end{figure}

In both the aggregate plot across all retailers (Fig.~\ref{fig:result_allbulk_ratio_location}) as well as in the specific 
retailers of Fig.~\ref{fig:result_ratio_location}, Finland appears to be getting consistently the higher prices among other locations. 
For this reason, we are tempted to examine whether this is indeed true across each and every retailer in the crawled dataset. 
For this reason we plot in Fig.~\ref{fig:result_Finland_ratio_location} the ratio between the price in Finland and 
the minimum price across all locations, for all the retailers of crawled. The results indicate that Finland is almost never the cheaper location (exceptions with \texttt{mauijim.com} and \texttt{tuscanyleather.it}).

\subsection{Personal information}
\label{subsec:personal_info}

In order to check if the personal information of users plays a role in price variations,
we first train personas as described in an earlier paper~\cite{hotnetspd}; we use an affluent and
a budget conscious persona. We check for prices of different products at these specific
e-retailers, taking measurements while keeping the location and time fixed, but we find \emph{no} price 
differences. 

\begin{figure}[tb]
\centering
\includegraphics[clip=true,trim=0cm 1.2cm 0cm 2cm,width=9cm]{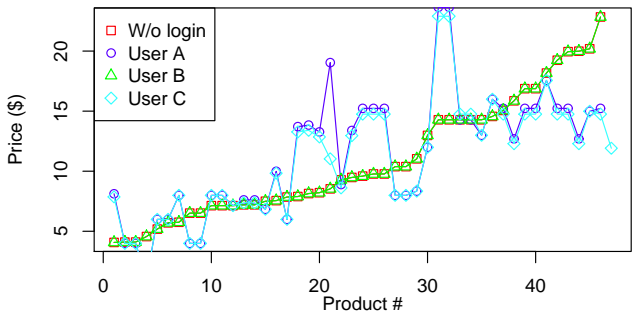}
\caption{The impact of login on the price of Kindle ebooks at www.amazon.com}
\label{fig:result_bulk_ratio_location}
\end{figure}

We do, however find some price variations for Kindle ebooks on \texttt{www.amazon.com}, depending
on if the user is logged in to the site or not. We present our results of collecting prices for
three users with different profiles and compare that against the price observed when there is no login. 
Our measurements are conducted at the same time and from the same location, and are plotted in 
Fig.~\ref{fig:result_bulk_ratio_location}. We note price variations for the same product and it would
appear there is little correlation to being logged in or not. There has been anecdotal evidence
about \texttt{amazon.com} varying prices dynamically in the past~\cite{wired-pd}, but for us to dig deeper
for reasons is currently beyond the scope of this paper. 

As a first step towards understanding the mechanism behind varying prices and
the parties that can possibly enable this, we investigate the frequency of third parties 
that are present on the retailers we study. It would appear that Google is present on most e-retailers with their analytics (95\%) and doubleclick (65\%) domains. Social networks have also significant presence on the retailers' sites through their widgets: Facebook (80\%), Pinterest (45\%), and Twitter (40\%). While we do \emph{not} see browsing history leading to price variations, it would be relatively easy for popular third parties to assist in price variations, fueled by the information they collect
across the web. We leave this to future work.

\section{Related work}
Price discrimination is as old as retail itself~\cite{mckenzie2008popcorn} but online price discrimination is a fairly new phenomenon. To the 
best of our knowledge one of the first to conjecture the rise of online price discrimination driven by large scale collection
of personal information was A. Odlyzko~\cite{Odlyzko2003}. 
The closest related work to the current paper is our previous article~\cite{hotnetspd}. In that paper we selected 
a set of popular retailers and observed variations of price based on location for two categories -- ebooks and office equipment 
and depending on the referring URI. 
In order to scale up the search process for price discrimination, we have turned to crowdsourcing that
enables us to efficiently study different retailers, leading to more instances of price variations, including in niche retailers and
different categories like clothing, hotels, automobiles, department stores, and photography equipment 
than reported earlier~\cite{hotnetspd}. 
Beyond price discrimination, personalization of the web using 
personal information of users is an active
area of research with the study of the filter bubble effect~\cite{Hannak2013}. We are interested
in the economic implications of personalization -- price discrimination on e-commerce 
domains. 

\section{Conclusions and Future Work}

In this paper we have analyzed the frequency and the magnitude of price variation observed in a crowdsourced and a more systematic crawled dataset. Our study makes a significant step compared with previous point results that we showed earlier on. Still, the examined retailers and products are a tiny drop in the ocean of today's e-commerce world. The resuts however are repeatable. Our intention is to keep collecting data and update the current picture that we have on the topic. This will hopefully occur once (and if) the user base of \emph{\$heriff} grows. In addition to scaling up the search for price discrimination it would be desirable if we could attribute the observed prices with the personal information of a user (\eg, web-sites visited, purchases performed, \etc). This is clearly an area that requires much more work.

\bibliographystyle{plain}
{\scriptsize \bibliography{conext}}

\end{document}